\documentstyle[12pt]{article}
\def\theequation{\arabic{section}.\arabic{equation}}
\begin{document}
\renewcommand{\thefootnote}{\fnsymbol{footnote}}

\renewcommand{\theequation}{\thesection.\arabic{equation}}

\title{The $N=2$ Supersymmetric Heavenly Equation\\ and Its Super-Hydrodynamical Reduction}

\author{Ziemowit Popowicz${}^a$\thanks{{\em e-mail: ziemek@ift.uni.wroc.pl}}
 and Francesco Toppan${}^b$\thanks{{\em e-mail: toppan@cbpf.br}}
\\ \\
${}^a${\it Institute for Theoretical Physics, University of
Wroc{\l}aw,}
\\ {\it 50-204 Wroc{\l}aw, pl. Maxa Borna 9, Poland}\\
 \\ ${}^b${\it CBPF, CCP, Rua Dr.}
{\it Xavier Sigaud 150,}
 \\ {\it cep 22290-180 Rio de Janeiro (RJ), Brazil}
}

\maketitle

\begin{abstract}
Manifest $N=2$ supersymmetric Toda systems are constructed from
the $sl(n,n+1)$ superalgebras by taking into account their complex
structure. In the $n\rightarrow \infty$ continuum limit an $N=2$
extension of the $(2+1)$-dimensional heavenly equation is
obtained. The integrability is guaranteed by the existence of a
supersymmetric Lax pair. We further analyze the properties of the
$(1+1)$-dimensionally reduced system. Its bosonic sector is of
hydrodynamical type. This is not the case for the whole
supersymmetric system which, however, is super-hydrodynamical when
properly expressed in terms of a supergeometry involving
superfields and fermionic derivatives.
\end{abstract}

\thispagestyle{empty} \vfill{CBPF-NF-003/03}

\section{Introduction}

The so-called ``Heavenly Equation" in $(1+2)$ dimensions was at
first introduced to describe solutions of the complexified
Einstein equations \cite{ple}. It has received a lot of attention
\cite{sav} for its remarkable integrability properties. Even if,
to our knowledge, no attempt so far in the literature has been
made to analyze the corresponding situation in the case of
supergravity, the supersymmetric heavenly equation has been
introduced and discussed in the context of superintegrable models.
It appears \cite{saso} as the continuum limit of a system of
equations known as ``supersymmetric Toda lattice hierarchy",
introduced in \cite{opeh}. Such system of equations and their
hidden $N=2$ supersymmetric structures have been vastly
investigated \cite{leso}. On the other hand, it has been recently
pointed out that the dimensional reduction of the
$(1+2)$-dimensional heavenly equation to a $(1+1)$-dimensional
system is related to Benney-like integrable hierarchies of
hydrodynamical equations \cite{cdp}. Another very recent reference
discussing the hydrodynamical properties of the reduced heavenly
equation is \cite{fp}.
\par In this paper we introduce the $N=2$ supersymmetric heavenly
equation, clarifying its integrability properties and its
algebraic arising from the $n\rightarrow \infty$ limit of a class
of $N=2$ supersymmetric Toda equations. Further, we investigate
its $(1+1)$ dimensional reduction which provides the
supersymmetric hydrodynamical equations extending the Benney-like
hierarchy of reference \cite{cdp}. The integrability properties of
such systems are induced by the integrability property of the
$N=2$ superheavenly equation and its Lax representation discussed
below. This Lax representation is not in the form of
supersymmetric dispersionless Lax operator, as we discuss at
length later.
\par
The plan of the paper is as follows, in the next section we
introduce the discretized $N=2$ Toda lattice hierarchy (as well as
its continuum limit, the $N=2$ Superheavenly Equation), as a
 supersymmetric Toda system based on the $sl(n|n+1)$ superalgebra.
 It is worth to remark that these are the same superalgebras originally
 employed in the construction of \cite{saso}.
 On the other hand, the superalgebras of the $sl(n|n+1)$ series
 admit a complex structure, allowing the
 construction of superToda systems based on $N=2$ superfields,
 according to the scheme of \cite{ivto}. For this reason, the next
 section results provide a generalization of those of
 \cite{saso}. The supersymmetric Lax pairs providing integrability
 of
 the systems (for any given $n$ and in the limit $n\rightarrow \infty$) are explicitly
 constructed. They provide a
 dynamical formulation in the $x_\pm$ plane (without involving the
 extra-time direction $\tau$ of the superheavenly equation).
 In the following, we investigate the dimensional reduction of both
 the discretized and continuum systems from $(1+2)$ to $(1+1)$
 dimensions. We obtain a supersymmetric system of equations with
 an interesting and subtle property. Unlike its purely bosonic subsector,
the whole system involving fermions is not of hydrodynamical type.
However, the same system, once expressed in terms of a
supergeometry involving superfields and fermionic derivatives,
satisfies a graded extension of the hydrodynamical property, which
can be naturally named of ``super-hydrodynamical type of
equation". With this expression we mean that there is a way to
recast the given ($1+1$)-dimensional supersymmetric equations into
a system of non-linear equations for superfields which involves
only first-order derivations w.r.t. the supersymmetric fermionic
derivatives. This super-hydrodynamical system furnishes the
supersymmetrization of the system introduced in \cite{cdp}.\par

\section{The $N=2$ Superheavenly equation.}

The construction of a continuum limit (for $n\rightarrow\infty$)
of a discretized superToda system requires a presentation of the
system in terms of a specific Cartan matrix. The symmetric choice
in \cite{kac} for the Cartan matrix of the superalgebra
$sl(n|n+1)$ does not allow to do so. On the other hand \cite{fss},
the Cartan matrix $a_{ij}$ of $sl(n|n+1)$ can be chosen to be
antisymmetric with the only non-vanishing entries given by $a_{ij}
=\delta_{i,i+1}-\delta_{i,i-1}$.\par The Cartan generators $H_i$
and the fermionic simple roots $F_{\pm i}$ satisfy
\begin{eqnarray}
\relax [H_i, F_{\pm j}] &=& \pm a_{ij}F_{\pm j}, \nonumber\\
\{F_i, F_{-j} \} &=& \delta_{ij}.
\end{eqnarray}
The continuum limit of the \cite{saso} construction could have
been performed for any superalgebra admitting an $n\rightarrow
\infty$ limit, such as $sl(n|n)$, etc. On the other hand, the
superalgebras of the series $sl(n|n+1)$ are special because they
admit a complex structure and therefore the possibility of
defining an $N=2$ manifestly supersymmetric Toda system, following
the prescription of \cite{ivto}. This is the content of the
present section.\par At first we introduce the $N=2$ fermionic
derivatives $D_\pm$, ${\overline D}_\pm$, acting on the $x_\pm$
$2D$ spacetime ($\theta_\pm$ and ${\overline\theta}_\pm$ are
Grassmann coordinates). The $2D$ spacetime can be either Euclidean
($x_\pm = x\pm t$) or Minkowskian ($x_\pm = x\pm i t$).\par We
have
\begin{eqnarray}
D_\pm &=& \frac{\partial}{\partial\theta_\pm}
-i{\overline\theta}_\pm \partial_\pm,\nonumber\\ {\overline D}_\pm
&=&
-\frac{\partial}{\partial{\overline\theta}_\pm}+i\theta_{\pm}\partial_\pm
.
\end{eqnarray}
They satisfy the anticommutator algebra
\begin{eqnarray}
\{ D_\pm, {\overline D}_\pm \}&=& 2i\partial_\pm
\end{eqnarray}
and are vanishing otherwise.\par Chiral ($\Phi$) and antichiral
($\overline \Phi$) $N=2$ superfields are respectively constrained
to fulfill the conditions
\begin{eqnarray}
{\overline D}_\pm {\Phi}&=& 0, \nonumber\\ D_\pm {\overline \Phi}
&=&0.
\end{eqnarray}
Accordingly, a generic chiral superfield ${\Phi}$ is expanded in
its bosonic ${\varphi}$, ${F}$ (the latter auxiliary) and
fermionic component fields (${\psi}_{+},{\psi}_{-}$) as
\begin{eqnarray}
{\Phi }({\hat x}_\pm, { \theta}_\pm) &=& {\varphi} +{\theta}_{+}
{\psi}_{+} +{\theta}_{-}\psi_{-} + {\theta}_+{\theta}_-{ F},
\end{eqnarray}
with ${\varphi}$, ${ \psi}_\pm$ and ${ F}$ evaluated in ${\hat
x}_\pm = x_\pm+i{\overline\theta}_\pm\theta_\pm$. \par Similarly,
the antichiral superfield ${{\overline\Phi}}$ is expanded as
\begin{eqnarray}
{\overline\Phi }({\overline x}_\pm, {\overline \theta}_\pm) &=&
{\overline\varphi} +{\overline\theta}_{+} {\overline\psi}_{+}+
 {\overline\theta}_{-}{\overline\psi}_{-}
+{\overline\theta}_+{\overline\theta}_-{\overline F},
\end{eqnarray}
with all component fields evaluated in ${\overline x}_\pm = x_\pm
-i{\overline \theta}_\pm \theta_\pm$.\par Due to the complex
structure of $sl(n|n+1)$, its Cartan and its simple (positive and
negative) root sector can be split into its conjugated parts
\begin{eqnarray}&
\begin{array}{ll}
  {\cal H} \equiv \{ H_{2k-1}\}, & \quad {\overline {\cal
H}}\equiv \{ H_{2k}\}, \\
  {\cal F}_+ \equiv\{ F_{2k-1}\}, & \quad {\cal F}_- \equiv \{ F_{-(2k-1)}\},\\
  {\overline {\cal F}_+} \equiv\{
F_{-{2k}} \}, &\quad {\overline{\cal F}_-} \equiv \{ F_{2k}\},
\end{array}&
\end{eqnarray}
for $k=1,2,\ldots, n$ .

Following \cite{ivto}, we can introduce the $sl(n|n+1)$ $N=2$
superToda dynamics, defined for the Cartan-valued chiral (${\bf
\Phi}$) and antichiral (${\bf {\overline\Phi}}$ ) $N=2$
superfields, \begin{eqnarray} {\bf \Phi} &=&
\sum_{k=1}^n\Phi_{k}H_{2k-1},\nonumber\\ {\bf {\overline\Phi}} &=&
\sum_{k=1}^n{\overline\Phi}_k H_{2k},
\end{eqnarray}
through the Lax operators ${\cal L}_\pm$ and ${\overline{\cal
L}_\pm}$, given by
\begin{eqnarray}
{\cal L}_+ &=& D_+{\bf \Phi} + e^{\bf{\overline\Phi}} {{F}_+}
e^{-\bf{\overline{\Phi}}},\nonumber\\ {\cal L}_- &=&- {F}_-
\end{eqnarray}
and \begin{eqnarray} {\overline{\cal L}}_+ &=& {\overline
F}_+,\nonumber\\ {\overline {\cal L}}_- &=& {\overline D}_-{\bf
{\overline \Phi}} + e^{\bf{\Phi}} {\overline F}_-e^{-\bf{\Phi}},
\end{eqnarray}
where
\begin{eqnarray}&
\begin{array}{ll}
  F_+ = \sum_k F_{2k-1}, & \quad  F_- =\sum_k F_{-(2k-1)},\\
  {\overline F}_+ = \sum_k F_{-(2k)},& \quad
  {\overline F}_- = \sum_k F_{2k},
\end{array}
&
\end{eqnarray}
(as before, the sum is over the positive integers up to $n$).\par
Explicitly, we have
\begin{eqnarray}
{\cal L}_+ &=& \sum_k (D_+\Phi_{k} H_{2k-1} +  e^{{\overline
\Phi}_{k-1}-{\overline \Phi}_{k}}F_{2k+1}),\nonumber\\ {\cal L}_-
&=& -\sum_kF_{-(2k-1)},\nonumber\\ {\overline{\cal L}}_+ &=&
\sum_k F_{-2k},\nonumber\\ {\overline{\cal L}}_- &=&
\sum_k({\overline D}_-{\overline \Phi}_k H_{2k} +
e^{\Phi_k-\Phi_{k+1}}F_{2k}),\nonumber\\ &&
\end{eqnarray}
Please notice that here and in the following we have formally set
${\overline\Phi}_0\equiv 0$.
\par The zero-curvature
equations are given by
\begin{eqnarray}
D_+ {\cal L}_- +D_-{\cal L}_+ +\{{\cal L}_+,{\cal L}_-\} &=&
0,\nonumber\\ {\overline D}_+{\overline {\cal L}}_- +{\overline
D}_-{\overline{\cal L}}_+ + \{{\overline{\cal
L}}_+,{\overline{\cal L}}_-\} &=& 0,
\end{eqnarray}
so that the following set of equations for the constrained
(anti)chiral $N=2$ superfields is obtained
\begin{eqnarray}
 D_+D_- \Phi_k &=& -e^{  {\overline
\Phi}_{k-1}-{{\overline\Phi}_k}  },\nonumber\\  {\overline
D}_+{\overline D}_-{\overline\Phi}_k &=& -e^{\Phi_k-\Phi_{k+1}},
\end{eqnarray}
for the positive integers $k=1,2,\ldots , n$. \par By setting,
\begin{eqnarray}
B_k = \Phi_k-\Phi_{k+1},&\quad& {\overline B}_k =
{\overline\Phi}_k -{\overline\Phi}_{k+1},
\end{eqnarray}
we get the two systems of equations
\begin{eqnarray}
D_+D_- B_k = e^{{\overline B}_k} -e^{{\overline B}_{k-1}},&\quad&
{\overline D}_+{\overline D}_- {\overline B}_k = e^{B_{k+1}} -
e^{B_{k}},
\end{eqnarray}
for $k=1,2,\ldots, n$. \par By identifying $k$ as a discretized
extra time-like variable $\tau$ we obtain, in the continuum limit
for $n\rightarrow \infty$,
\begin{eqnarray} \label{2shev}
D_+D_- {B} =
\partial_\tau e^{{\overline B}},&\quad & {\overline
D}_+{\overline D}_- {{\overline B}}=
\partial_\tau e^{B},
\end{eqnarray}
which corresponds to the $N=2$ extension of the superheavenly
equation.\par Indeed, the presence in the previous equations of
the first derivative in $\tau$ is an artifact of the $N=2$
superfield formalism. Once solved the equations at the level of
the component fields and eliminated the auxiliary fields in terms
of the equations of motion, we are left with a system of
second-order equations. \par We have, in terms of the component
fields,
\begin{eqnarray}
B&=& \Big (1+i{\overline\theta}_+\theta_+\partial_{+}
+i{\overline\theta}_-\theta_-\partial_{-}-{\overline\theta}_+\theta_+
{\overline\theta}_-\theta_-\partial_+\partial_-\big ) {\cal C},
\nonumber \\ {\overline B}&=& \Big
(1-i{\overline\theta}_+\theta_+\partial_+
-i{\overline\theta}_-\theta_-\partial_- -{\overline\theta}_+
\theta_+{\overline\theta}_-\theta_-\partial_+\partial_-\Big
){\overline {\cal C}},
\end{eqnarray}
where
\begin{eqnarray}
{\cal C} &=&  \Big (b + \theta_+\psi_{+}+\theta_-\psi_{-}  +
\theta_+\theta_-a\Big ),\nonumber \\ {\overline {\cal C}}&=& \Big
({\overline
b}+{\overline\theta}_+{\overline\psi}_++{\overline\theta}_
-{\overline\psi}_-+{\overline\theta}_+{\overline\theta}_-{\overline
a} \Big ),
\end{eqnarray}
with $a$, ${\overline a}$ bosonic auxiliary fields. All component
fields are evaluated in $x_{\pm}$ only.\par The equations of
motion of the $N=2$ superheavenly equation are explicitly given in
components by
\begin{eqnarray}
-a &=& (e^{\overline b})_\tau ,\nonumber\\
2i\partial_-\psi_+&=&({\overline \psi}_-e^{\overline b})_\tau ,
\nonumber\\ -2 i\partial_+\psi_- &=& ({\overline
\psi}_+e^{\overline b})_\tau ,\nonumber\\
-4\partial_+\partial_-b&=& ({\overline a}e^{\overline
b}-{\overline \psi}_+{\overline\psi}_-e^{\overline b})_\tau ,
\end{eqnarray}
and
\begin{eqnarray}
- {\overline a}&=&(e^b)_\tau,\nonumber\\
2i\partial_-{\overline\psi}_+&=&(\psi_-e^b)_\tau,\nonumber\\ -2
i\partial_+{\overline\psi}_-&=& (\psi_+e^b)_\tau,\nonumber\\
-4\partial_+\partial_-{\overline b}&=&
(ae^b-\psi_+\psi_-e^b)_\tau.
\end{eqnarray}
Eliminating the auxiliary fields we obtain the systems
\begin{eqnarray} \label{shev1}
 2i\partial_-\psi_+&=&({\overline \psi}_-e^{\overline b})_\tau ,\nonumber\\
 -2 i\partial_+\psi_- &=& ({\overline
\psi}_+e^{\overline b})_\tau ,\nonumber\\ 4\partial_+\partial_-b
&=& \Big ( (e^{b})_{\tau} e^{\overline b}+({\overline
\psi}_+{\overline\psi}_-e^{\overline b} \Big )_{\tau} ,
\end{eqnarray}
and
\begin{eqnarray} \label{shev2}
 2i\partial_-{\overline\psi}_+&=&(\psi_-e^b)_\tau,\nonumber\\ -2
i\partial_+{\overline\psi}_-&=&  (\psi_+e^b)_{\tau},\nonumber\\
4\partial_+\partial_-{\overline b} &=& \Big (
(e^{\overline b})_{\tau}e^{b} + (\psi_+\psi_-e^b\Big )_\tau.
\end{eqnarray}
The bosonic component fields $b$, ${\overline b}$, as well as the
fermionic ones $\psi_\pm$, ${\overline\psi}_\pm$, are all
independent. The equations (\ref{shev1}, \ref{shev2}) are a
manifestly $N=2$ supersymmetric extension of the system introduced
in \cite{saso}.

\section{Super-hydrodynamical reductions of the superheavenly equation.}

The equations (\ref{shev1}) and (\ref{shev2}) correspond to a
$(1+2)$-dimensional system, manifestly relativistic and $N=2$
supersymmetric w.r.t. the two-dimensional subspace spanned by the
$x_\pm$ coordinates, while possessing an extra bosonic time-like
dimension denoted as $\tau$. A very interesting example of
integrable system in $1+1$-dimension which is currently intensely
investigated, see e.g. \cite{{cdp},{fp}}, can be recovered by
applying a dimensional-reduction to, let's say, the bosonic sector
of the $(1+2)$-dimensional heavenly equation. We can refer to such
a system, perhaps a bit pedantically, as the $(1+1)$-dimensionally
reduced heavenly equation. It can be obtained by setting equal to
zero the fermionic variables $\psi_\pm, {\overline \psi}_\pm\equiv
0 $, while $b$, ${\overline b}$, can be consistently constrained
as ${\overline b}=b$. The $x_\pm$ coordinates are identified, i.e.
$x_+=x_-=x$.\par The resulting equation, by changing the
normalizations (setting $f=2b$ and $t=2\tau$) can be conveniently
written as
\begin{equation} \label{zhev}
f_{tt} = (e^f)_{xx}
\end{equation}
This equation has recently received a lot of attention in the
literature. It is an example of a completely integrable,
hydrodynamical-type equation. It admits a multihamiltonian
structure and possesses an infinite number of conserved charges in
involution. It has been recently shown that it can be recovered
via a dispersionless Lax representation given by
\begin{equation}
L:=p^{-1}\Big ( 1+gp^2+hp^4 \Big )^{\frac{1}{2}},
\end{equation}
with $f,g$ functions of $x,t$, while $p$ is the classical
momentum. The equations of motion are read from
\begin{equation} \label{poiss}
\frac{\partial L}{\partial t} = \frac{p}{2} \Big \{ L^{2}_{\leq
0}, L \Big \}
\end{equation}
where $ \{\ast , \ast \}$ denotes the usual Poisson brackets and
$L^{2}_{\leq 0}$, defined in \cite{cdp}, is explicitly given by
$L^{2}_{\leq 0} = p^{-2}+g$.\par The equation (\ref{poiss}) leads
to
\begin{eqnarray} \label{2comp}
\frac{\partial h}{\partial t} &=& hg_x \\ \nonumber \frac{\partial
g}{\partial t} &=& h_x \label{hydro}
\end{eqnarray}
The equation (\ref{zhev}) is recovered after eliminating $g$ from
the previous system and setting $h=e^{f}$. In the (\ref{hydro})
form, the ``$(1+1)$-dimensionally reduced heavenly equation" is
shown to be a hydrodynamical type of equation.\par An interesting
question, whose solution as we will see is non-trivial, is whether
the $(1+1)$-dimensional reduction (for $x_+=x_-=x$) of the full
$N=2$ supersymmetric heavenly equation is also of hydrodynamical
type. The answer is subtle. The introduction of the fermionic
fields $\psi_\pm$, ${\overline\psi}_\pm$, whose equations of
motion are of first order in the extra-bosonic time, does not
allow to represent the dimensionally reduced system from
(\ref{shev1}, \ref{shev2}) in hydrodynamical form, due to a
mismatch with the second-order-equation satisfied by the bosonic
fields $b$, ${ \overline b}$. On the other hand, it is quite
natural to expect that important properties are not spoiled by the
supersymmetrization. This is indeed the case with the
hydrodynamical reduction, when properly understood. The key issue
here is the fact that the nice features of the supersymmetry are
grasped when inserted in the proper context of a supergeometry,
which must be expressed through the use of a superfield formalism.
It is in this framework that a super-hydrodynamical reduction of
the dimensionally reduced system from (\ref{shev1}, \ref{shev2})
becomes possible. Indeed, for $x_+=x_-=x$, we can write down the
(\ref{2shev}) system through the set of, first-order in the
fermionic derivatives, {\em superfield} equations
\begin{eqnarray}\label {comp1}
{\cal D}_{-} {\cal B} & = &  N_{\tau}, \\ \nonumber {\cal D}_{+} N
& = & e^{\overline B}
\end{eqnarray}
and
\begin{eqnarray}\label {comp2}
{\overline {\cal D}}_{-} {\overline {\cal B}} & = &  {\overline
N}_{\tau}, \\ \nonumber {\overline {\cal D}}_{+} {\overline N} & =
& e^B
\end{eqnarray}
with $N$ and ${\overline N}$ subsidiary  fermionic superfields .
Taking into account the following expansions
\begin{eqnarray}
e^{\overline B}&=&(1-i{\overline\theta}_+\theta_+\partial_+
-i{\overline\theta}_-\theta_-\partial_-
-{\overline\theta}_+\theta_+{\overline\theta}_-\theta_-\partial_+\partial_-)
 e^{\overline {\cal C} } = \nonumber \\
 && {\cal D}_{+}\theta_{+}
 \Big (1-i{\overline {\theta_{-}}}\theta_{-}\partial_-\Big )e^{\overline {\cal C}}\nonumber \\
e^{B}&=&(1+i{\overline\theta}_+\theta_+\partial_+
+i{\overline\theta}_-\theta_-\partial_-
-{\overline\theta}_+\theta_+{\overline\theta}_-\theta_-\partial_+\partial_-)
 e^{{\cal C} } = \nonumber \\
&& -
{\overline {\cal D}}_{+}{\overline\theta}_+
\Big ( 1+i{\overline\theta_-}\theta_-\partial_-\Big )e^{\cal C}
\end{eqnarray}
we easily obtain the following solutions for $N$, ${\overline N}$
\begin{eqnarray}
N &=&
 \Big (\theta_{+}-i\theta_{+}
 {\overline\theta}_{-}\theta_{-}\partial_{-} \Big ) e^{\overline {\cal C}}
 + {\cal D}_{+}\Omega_1 \nonumber \\
{\overline N} &=&
 - \Big ({\overline\theta}_{+}+
 i{\overline\theta}_{+}{\overline\theta}_{-}\theta_{-}\partial_{-} \Big )
 e^{\cal C} + {\overline {\cal D}}_{+}{\overline\Omega}_1
 \label{nnbar}
\end{eqnarray}
in terms of arbitrary $\Omega_1$, $\Omega_2$ bosonic
superfunctions. \par If we set, as we are free to choose,
\begin{eqnarray}
\Omega_1 &=& \theta_{+} \Big
(\Psi_{-}-2i{\overline\theta}_{-}b_{-} -i{\overline\theta}_{-}
\theta_{-} \Psi_{-,-}\Big ) \nonumber  \\ {\overline\Omega}_1 &=&
{\overline\theta_{+}} \Big
({\overline\Psi}_{-}-2i\theta_{-}{\overline b}_{-}
+i{\overline\theta}_{-}\theta_{-} {\overline\Psi}_{-,-}\Big )
\end{eqnarray}
and substitute the values of $N$, ${\overline N}$ given in
(\ref{nnbar}) back to (\ref{comp1}) and (\ref{comp2}), we obtain
the equivalence of this system of equations w.r.t. the $x_+=x_-=x$
dimensional reduction of the (\ref{shev1}) and (\ref{shev2})
equations. This proves the existence of a super-hydrodynamical
reduction expressed in a superfield formalism. If we express in
terms of the component fields this set of equations, as mentioned
before, we are not obtaining a hydrodynamical type equation. This
fact should not be regarded as a vicious feature of our system,
but rather as a virtue of the supersymmetry. In several examples,
this is one, the introduction of the super-formalism allows
extending both the properties of the systems and the techniques
used to investigate them, to cases for which the ordinary methods
are of no help. In a somehow related area we can cite, e.g., the
derivation of bosonic integrable hierarchies associated with the
bosonic sector of super-Lie algebras \cite{anp}. They are outside
the classification based on and cannot be produced from ordinary
Lie algebras alone.\par Let us finally discuss the restriction
from $N=2\rightarrow N=1$, i.e. down to the $N=1$ supersymmetry.
It is recovered from setting
\begin{eqnarray}
{\overline\theta}_+=-\theta_{+},
&\quad&{\overline\theta}_-=-\theta_-, \nonumber\\ {\overline {\cal
D}}_+= -{\cal D}_+, &\quad& {\overline {\cal D}}_-=-{\cal D}_-.
\end{eqnarray}

The equations (\ref{shev1}) and (\ref{shev2}) now read
\begin{eqnarray}\label{sushev}
{\cal D}_+{\cal D}_-{\cal C} = \partial_{\tau}e^{\overline{\cal
C}}, &\quad& {\cal D}_+{\cal D}_-{\overline {\cal C}} =
\partial_{\tau}e^{\cal C},
\end{eqnarray}
with
\begin{equation}
D_\pm = \frac{\partial}{\partial\theta_\pm}
+i\theta_\pm \partial_\pm
\end{equation}
The equations in (\ref{sushev}) have a similar structure in
components as the equations (\ref{shev1}) and (\ref{shev2}). We
have indeed
\begin{eqnarray}
 \partial_-\psi_+&=&i({\overline \psi}_-e^{\overline b})_\tau ,\nonumber\\
 \partial_+\psi_- &=& -i({\overline\psi}_+e^{\overline b})_\tau ,\nonumber\\
\partial_+\partial_-b &=& \Big ( (e^{b})_{\tau} e^{\overline b}+({\overline
\psi}_+{\overline\psi}_-e^{\overline b} \Big )_{\tau} ,
\end{eqnarray}
and
\begin{eqnarray}
 \partial_-{\overline\psi}_+&=& i(\psi_-e^b)_\tau,\nonumber\\
 \partial_+{\overline\psi}_-&=& -i(\psi_+e^b)_{\tau},\nonumber\\
 \partial_+\partial_-{\overline b} &=& \Big (
(e^{\overline b})_{\tau}e^{b} + (\psi_+\psi_-e^b\Big )_\tau.
\end{eqnarray}

If we further dimensionally reduce to $(1+1)$-dimension our $N=1$
system, by constraining the variables $x_{\pm}$ ($x_+=-x_-=it$)
and setting ${\overline {\cal C}}={\cal C}$, we obtain the system
\begin{eqnarray}
\partial_t\psi_+ &=& \partial_{\tau}\Big ( e^b\psi_- \Big ) \nonumber \\
\partial_t\psi_- &=& \partial_{\tau} \Big ( e^b\psi_+ \Big ) \nonumber \\
\partial_t^2 b &=& \partial_{\tau} \Big ( \frac{1}{2} (e^{2b})_{\tau} +e^b\psi_+\psi_-\Big
).
\end{eqnarray}
As its $N=2$ counterpart, this system is not of hydrodynamical
type. However, it can be easily shown, with a straightforward
modification of the procedure previously discussed for the $N=2$
case, to be super-hydrodynamical when expressed in terms of
superfields and superderivatives.

\section{Conclusions.}

In this paper we have at first introduced manifest $N=2$
supersymmetric discrete Toda equations in association with the
complex-structure superalgebras of the $sl(n,n+1)$ series. We
proved that in the $n\rightarrow \infty$ limit the discrete
variable can be regarded as a continuum extra bosonic time $\tau$.
The corresponding system is a manifestly $N=2$ supersymmetric
extension of the standard heavenly equation in $(2+1)$ dimensions,
the supersymmetry being associated with the plan spanned by the
bosonic $x_\pm$ coordinates and their fermionic counterparts,
which are naturally expressed in an $N=2$ superspace formalism. A
supersymmetric Lax-pair, proving the integrability of the system,
was constructed. It expresses the dynamics w.r.t. the $\pm$
superspace coordinates ($\tau$ in this respect can be considered
either as an auxiliary parameter or as the discrete label of the
original formulation).\par We further investigated the dimensional
reduction of the above systems from the $(2+1)$-dimensional to the
$(1+1)$-dimensional case. In the purely bosonic case the reduced
systems of equations are of a special type, they are
hydrodynamical systems of non-linear equations. The supersymmetric
case is much subtler. This system is not of hydrodynamical type
since the presence of the fermions spoils this property. However,
when properly expressed in terms of a supergeometry (essentially,
superfields and fermionic derivatives) is of super-hydrodynamical
type (the precise meaning of this term has been explained in the
Introduction). More than a vice, this can be considered as a
virtue of the supersymmetry. It allows extending the notion of
hydrodynamical equations beyond the realm of the systems ordinary
allowed.\par Finally, it is worth mentioning a formal, however
challenging problem, concerning the supersymmetrization. While the
integrability of the systems under consideration is automatically
guaranteed by the given supersymmetric and relativistic Lax-pairs
in the $\pm$ plane mentioned above, one can wonder whether a
dispersionless Lax operator, directly expressing the ${\tau} $
dynamics (i.e., in terms of the bosonic extra-time) could be
found. The answer is indeed positive for the purely bosonic
sector. In the supersymmetric case, however, no closed
dispersionless Lax operator is known at present. It is repeated
here an analogous situation already encountered for the polytropic
gas systems \cite{dp}. The problem of constructing supersymmetric
dispersionless Lax operators for these related systems is still
open.
\\
{\quad}\par

 {\large{\bf Acknowledgments.}}
~{\quad}\\{\quad}\par Z.P. is grateful for the hospitality at
CBPF, where this work was initiated, while F.T. is grateful for
the hospitality at the Institute of Theoretical Physics of the
University of Wroc{\l}aw, where the paper has been finished.

\end{document}